\documentclass[sn-mathphys,Numbered]{sn-jnl}

\usepackage{graphicx}%
\usepackage{multirow}%
\usepackage{amsmath,amssymb,amsfonts}%
\usepackage{amsthm}%
\usepackage{mathrsfs}%
\usepackage[title]{appendix}%
 \usepackage{colordvi}\usepackage[usenames]{color}
\usepackage{textcomp}%
\usepackage{manyfoot}%
\usepackage{booktabs}%
\usepackage{algorithm}%
\usepackage{algorithmicx}%
\usepackage{algpseudocode}%
\usepackage{listings}%

\theoremstyle{thmstyleone}%
\theoremstyle{thmstyletwo}%

\theoremstyle{thmstylethree}%

\newcommand{\etal}{{\it et al.}}
\newcommand{\cita}[1]{{\it``#1''\/}}

\begin{document}

\title[Article Title]{The story around the first 4n signal}


\author*[1]{\fnm{F.~Miguel} \sur{Marqu\'es}}\email{marques@lpccaen.in2p3.fr}

\affil*[1]{LPC Caen, ENSICAEN, CNRS/IN2P3, Universit\'e de Caen, Normandie Universit\'e, 14050 Caen, France}


\abstract{The GANIL campaign around the first 4n signal was very peculiar. The beginning and end were both dictated by unexpected events that, unfortunately, do not fit within the streamlined format of standard scientific publications.
 However, they illustrate many aspects of how basic research should work, or at least does work. Therefore, I take this opportunity to share them with those not involved in the campaign, hoping that they will offer a better perspective of that research in particular and of basic research in general. As a disclaimer, this is only a personal recollection of those events.}




\maketitle

\section{Introduction}

 This paper is not about physics.
 The physics of multineutron systems in general can be found in Ref.~\cite{Review}, and the case of the tetraneutron in particular with the recent 4n signal confirmation in Ref.~\cite{Nature}.
 This paper is about things that have little to do with physics. I have taken the opportunity to use the first 4n signal work \cite{GANIL} as an illustration of the `non-physics' matters that are never discussed in formal publications.
 Physics papers are expected to follow a pre-established format of linear success, and in practice they all do. We are supposed to find ideas, propose experiments, perform them, analyze the data, find new results, and finally reach conclusions and open perspectives.
 
 Everything else out of this well-defined path does not fit the publication format, and as a consequence the readers build themselves an ideal picture in which it does not happen.
 However, it is well-known that basic research is not made of linear successes. Most of the discoveries come from unexpected results, unplanned observations of past experiments, byproducts of unrelated analyses... And many experiments do not lead to the expected results, sometimes to any result at all.
 The story of unexpected results is often deformed so that it fits the linear success model of our books and journals, and null experiments are mostly ignored. Therefore, the real chain of events is cornered into some form of oral tradition that ends up fading away.
 
 Besides the fact that scientific papers do not reflect real-life research, this phenomenon has a very negative impact on young researchers. They believe that all results should be planned, all experiments should lead to results, and therefore they see null experiments as unforgivable failures. They do not feel the incentive for risk taking, an essential element of basic research, and their general practice becomes closer to that of applied research.
 The aim of this paper is to counter this impression. Basic researchers should be allowed to take risks and perform null experiments. The whole campaign around the first 4n signal is a good example of unexpected results and unintended failures, something you will never learn from the original publication.  
 
\section{New ideas}

 The idea of the first 4n signal originated during a random read at the library of LPC, early 2001. This library does no longer exist, as most of the former libraries of our laboratories, replaced by an online subscription to a list of selected journals that we can read at our office. However, at that time it was usual to leave the office once a week, go to the library, sit down and read the headlines of the latest journals. And one of those days I came across a strange paper \cite{Grater99}: \cita{Search for a bound trineutron...}.

 I was astonished. Why would someone search for something we all know does not exist? And then I went beyond the abstract and realized that it was not a one-shot nonsense try or byproduct. It was a well-thought experiment within a vast research program on multineutron systems, starting in the 1960s (later on I wrote a review of the program with Jaume Carbonell \cite{Review}).
 At that time we were studying light neutron-rich nuclei, but I did not know that other groups were searching for neutron-only nuclei, and I immediately made a link between both fields. How had they been searching for them? What were their limitations? Could we add something new to the search?
 
 I honestly doubt I would have been exposed to that work today. Our online access is often tailored so as to show us our field only, and thus avoid `wasting' our precious time on other fields. Accessing those works with the same computer with which we are almost in parallel working on our data does not help either. Exposure to different fields, and outside our usual working space, is the best way to find new ideas. Unfortunately libraries have become rare, but interdisciplinary conferences like the Few-Body Systems series represent still a unique opportunity for this cross-fertilization.

\section{Risky experiments}

 In short, multineutrons had been searched for using very low cross-section reactions, because creating a system of neutrons from nuclei that have similar numbers of neutrons and protons involves very low probabilities. Therefore, the absence of positive signals was generally attributed to the low probability of the probe and/or the consequently high relative background, not to the non-existence of multineutrons.
 It is easier to prove that something exists, through its observation, than that something does not exist.
 But even if we assumed that multineutrons did not exist, the question was: {\em if\/} they existed, could we use a higher-probability creation mechanism? And {\em if\/} we created them, how could we see them?
 
 The first question was straightforward. I said we were working on neutron-rich nuclei. If multineutrons existed, they could preexist inside those nuclei, like $\alpha$ particles do inside $\alpha$-decaying nuclei. For example, the $^8$He wave function could have some component of the type $^4$He+$^4$n. Moreover, neutron-rich nuclei are weakly bound, since they are more unstable than their natural isotopes, and the typical cross-sections to break them up are thus much higher.
 Therefore, breaking up neutron-rich nuclei should be better suited to the search for multineutrons.
 
 The second question was harder. We were detecting neutrons at GANIL, with energies from few to tens of MeV, using the scintillator modules of the DEMON array. The principle was the same that Chadwick used to discover the neutron \cite{Chadwick}: measure the recoil of a charged particle, in our energy regime a proton from the scintillator. A neutron, with a similar mass, cannot communicate more than its own velocity to the proton... but what if the `neutron' was heavier?
 We could tag reaction channels in which the projectile lost 4n and check for abnormally high proton recoil energies.

 We considered submitting a proposal of experiment. The physics case seemed strong, the new probe promising, and the signal would be clean. However, asking for beam time in search for something that should not exist is not an easy task. We were told that accelerators can eventually accept such proposals, but that they could not accept many and/or very risky experiments without a guaranteed result.
 As a disclaimer, we had proposed some of those already! We were asked to demonstrate first that the tool worked, namely that we could calibrate the proton recoil energy and that there would be no background in the expected signal region.

\section{The first signal}

 We had data from a previous campaign in 1997 focused on the properties of 2n halo nuclei ($^6$He, $^{11}$Li and $^{14}$Be) \cite{Haloes}. We used them to calibrate the proton recoil energy, compared it to the energy of the neutron (deduced from its time of flight from target to detector), and checked that the ratio $E_p/E_n$ was lower than 1. And it was indeed, in fact lower than about 1.4 due to the resolution of the detectors. Well, it {\em mostly\/} was, there were a few events (six) beyond 1.4, which we assumed to be a reasonable level of background ($\sim1\%$). As a curiosity, events with $E_p/E_n>1$ were until then traditionally rejected in the different analyses as impossible events.
 
 Most of the reactions led to the loss of 2n, the most likely process in the breakup of 2n halo nuclei. But those six events belonged {\em all\/} to a lower-probability reaction channel: $^{14}$Be leading to $^{10}$Be, the loss of 4n... It was May 17, 2001. Had we found a tetraneutron signal while preparing ourselves to search for it?
 The claim was too important, so we spent a lot of time in the search for alternative sources of these six events. We could not find any significant one. Instead of a more visible Letter, we chose to submit a longer paper to Physical Review C, with enough room for all the analysis details of the potential alternative scenarios.

 Our original title was \cita{On the detection of neutron clusters}. We had found a few events, but we wanted the paper to be mainly about our new proposal of multineutron probe, and then in its illustration discuss the possible origins of these events. However, the referees and editor were very enthusiastic and they decided to remove the first two words from the title, which became thus more categorical: \cita{Detection of neutron clusters} \cite{GANIL}. It appeared (unintentionally) on Fool's Day, April 1, 2002.

\section{Impact on the field}

 The first reaction was disbelief. Almost at the same time a theoretical paper \cite{3n} had shown that exact three-body calculations of the 3n system forbade bound or resonant states. A few months later another theoretical paper \cite{4n} concluded that the 4n system would not have bound states either, although using an approximated method (four-body exact calculations were not available yet). However, it mentioned a remote possibility of it having a broad, resonant state. Despite the non-exact character of these calculations, the community took both conclusions for granted: tetraneutron clusters did not exist; and maybe 4n resonances could.
 
 Another theoretical paper \cite{Carlos} argued that even a tetraneutron cluster would not be able to push a proton forward without breaking up in the process, due to its weak binding. We considered both hypothetical scenarios, the possible existence as a resonance and the likely breakup of a cluster, and demonstrated that they would have produced a similar signal through the detection of the `pieces' (several neutrons) inside the same detector module. These conclusions represented a strong reinforcement of the original signal, and as such we submitted them to the Rapid Communications section of Physical Review C. However, the referee dismissed them due to the \cita{absence of new data}... We will come back to this absence of new data in Section~\ref{s:NewData}.

 In fact we were too involved at that moment in the collection of new data, so much that we did not want to spend time arguing with the referee and we left our work available only through the ArXiV \cite{ArXiV}. Our conclusion was that the 4n system could have been responsible of the signal we had observed if it had a state with an energy between $-1$~MeV (the difference between the $^4$He and $^6$He thresholds in $^8$He) and some $+2$~MeV (the range in which the decay in flight would send several neutrons into one detector module). Other experimental colleagues searched for resonances (since clusters were said to be impossible) for a couple of years without success, and the field became calm again.

\section{Impact on the media}

 Due to the potential importance of the results, the CNRS published a press release that very soon attracted a lot of journalists. Most of them wrote short notes or articles on their own, some contacted us and asked for an interview, and some of those let us reread what they wrote to check possible misunderstandings.
 We were glad to see that many media were interested in our results, and that most of them made their best to remain faithful to our original publication and conclusions.
 The coverage was very wide, reaching the general public from science magazines to general newspapers. These are some excerpts I have kept: \bigskip \def\vof{\\[2mm]}
 
\begin{center}
 \cita{Atom: the pure nucleus, an object of the third kind} \\ {\sl L'Express} (French magazine) \vof
 \cita{A mysterious quadrineutron has probably been seen at GANIL} \\ {\sl Le Monde} (French journal) \vof
 \cita{A four-neutron nucleus is the only explanation for the results} \\ {\sl Science \& Vie} (Frech magazine) \vof
 \cita{A foretaste of tetra-neutrons} \\ {\sl Physics World} \vof
 \cita{Element zero. Also called tetraneutron because it would be composed of 4 neutrons} \\ {\sl Focus} (Italian magazine) \vof
 \cita{Element zero? Experiments have found a new type of matter:\,
 \setlength{\fboxsep}{0.5\fboxsep}\fbox{\rm$^4_0$Tn}\setlength{\fboxsep}{2\fboxsep}} \\ {\sl New Scientist}
\end{center} \bigskip

 A few of them, however, were deceiving, the {\sl New Scientist} treatment being the worst. We had insisted on caution and respect of our conclusions, asked to avoid exaggerated claims on the front cover, explained that we had not observed ``element zero'' (even if one accepted the concept, this tetraneutron would be just a heavier isotope of the neutron)... 
 The journalist agreed on all points, but then used the quote above as the huge headline of the front cover, even making up a new symbol for this new element!
 In the inner pages, the article \cita{Ghost in the atom} started with an eye-catching line, \cita{In a concrete bunker between Cherbourg and Paris...}, but then it even evoked \cita{some kind of experimental error}.
 {\sl Focus} also mentioned the new element and the possible error, but at least not in the front cover.
 
 Despite the bitter taste left by those few articles that distorted our actual result, we should not forget that outreach is very important in basic science, and in particular in nuclear physics. Jim Al-Khalili wrote an article on this particular case \cite{Jim}, and concluded that \cita{the publicity generated by the New Scientist article is priceless}. Many people, young and not, that had neglected nuclear physics as an old and boring field became attracted to it, no matter the inaccuracies or misleading shortcuts.
 One would wish to have absolute control on the way our results are presented to the general audience, but the fact is that we do not have it. Nevertheless, we should keep working on outreach, society must know that basic research is a unique strength and should be aware of what we do and how the process of discovery works in real life.
 
 Finally, this first 4n signal reached even Wikipedia! We do not know who created the ``Tetraneutron'' page, but checking the history it was done on October 5, 2002. Three years later, someone else updated the content following the theoretical work on the breakup of an hypothetical cluster \cite{Carlos}, saying that \cita{at least part of the original analysis was flawed}.
 A few months ago I edited the page (my only contribution) to correct at last that false claim: \cita{... but the suggestion was refuted\/ {\rm\cite{ArXiV}}}. The page is still growing with the most recent results.
 Wikipedia has become a very common source of information, and as such I welcome the existence of this page, but please keep in mind the anonymous character of contributors and that some things may not have been checked by specialists.

\section{The harsh reality} \label{s:NewData}

 The only publications from the 4n campaign at GANIL were Refs.~\cite{GANIL,ArXiV}, about the signal obtained with data from a previous experiment (E295, back in 1997). Therefore, the official version for the community was that no experimental attempt at confirmation was ever performed, not even a dedicated experiment (the signal had come from an unexpected byproduct). However, there were several experiments, in fact this was our main activity at GANIL during the next 5 years!
 Why was this activity unknown to the community? Why did people have the impression that we had done a one-shot try and then moved on to something else?

 The 4n signal from E295 was obtained during less than 2 days with a $^{14}$Be beam at an intensity of 130~pps (particles per second). After seeing our preliminary results, not yet published, the GANIL direction decided to reassign another experiment we were going to run in September 2001 (E378, on heavy helium isotopes) to the search for more tetraneutron events in the breakup of $^{14}$Be, expecting a beam more intense than in 1997. Unfortunately, several problems in the production of this beam led to only 20~pps, and in agreement with GANIL we decided to make the most of the remaining beam time by changing to another beam and physics case.
 
 In November 2002, after the publication of the 4n signal and the previous failure of beam production for E378, the GANIL direction decided to reassign once more the beam time of E378 to a different beam, $^8$He, easier to produce at higher intensities. All the setup was adapted to the search for the $^4$He+$^4$n channel, but some detector issues during the first day were followed by a strike action of the accelerator operators. Apparently they found the experiment important enough for the directors to listen to their claims, but they did not... The experiment was canceled.

 A few weeks later, in December 2002, we were finally ready to run the first dedicated official proposal for the search of multineutrons (E415), with a $^{14}$Be beam delivered by the newly commissioned separator LISE2000.
 We were supposed to run for 8 days with an intensity of 500~pps, which should allow us to confirm the 4n signal. Again, problems in the production of this beam led after several days to only 100~pps, moreover accompanied by a background of lighter particles at a rate of 10,000~pps!
 The online analysis showed that no signal could be expected under such unfavorable conditions, and we decided to change to an easier beam, $^{12}$Be, and search for the $\alpha$+$\alpha$+$^4$n channel. And again, the accelerator operators went on strike, for the same unresolved claims they had raised a month earlier.
 
 The DEMON campaigns at GANIL were a complex affair. Most of the detectors were not in-house. The charged-particle detector array belonged to the British CHARISSA collaboration and the neutron array to the Franco-Belgian DEMON collaboration. The campaigns at GANIL required thus a three-fold request preparation, to the CHARISSA, DEMON and GANIL committees. Once they all agreed and matched, the coming of the detectors to GANIL was organized, everything was set up during several months, if possible a series of experiments were scheduled and run in order to optimize the investment, and then everything was dismantled and moved away. The whole process involved the work of about a hundred people.
 
 The beam production issues and the strike actions were therefore particularly upsetting. When the strike was finally over the detectors had already left, and the net result was that the activity of so many people during so many months had led to literally nothing. Nevertheless, a few months later we were already working on all the necessary steps aiming at the next DEMON campaign at GANIL.
 Following the recent experience it had been clear that GANIL could not guarantee a clean beam of $^{14}$Be at more than 100~pps, which made us consider another avenue, the proton removal from a less exotic $^{15}$B beam. All the stars (detectors) aligned again in April 2006 for the second dedicated tetraneutron experiment (E483), in search this time for the reaction $^{15}$B$\,\rightarrow\,^{14}$Be$^*\rightarrow\,^4$n.
 
 Unfortunately, no significant signal was observed, although this did not disprove the first one because it was a very different reaction channel. In the original scenario we had supposed that a significant $^{10}$Be+$^4$n component could be present in $^{14}$Be ground state.
 Here, we were searching for such a configuration in the excited states (all unbound) of $^{14}$Be, and it turned out that we populated almost exclusively the first excited state, which decays into $^{12}$Be+2n and cannot emit 4n. More than 5 years had passed from the observation of the 4n signal with a very strong commitment of too many people, and we decided to stop `beating a dead horse'.
 At that time RIKEN (in Japan) was already delivering much more intense neutron-rich beams, for example their $^{14}$Be intensity was a hundred times more than the one we had had at GANIL (it is a thousand times more nowadays).
 
\section{Move to Japan}

 Therefore, a few years later we started to shift our experimental program on light neutron-rich nuclei to RIKEN, and in 2012 we did run our first campaign with the SAMURAI spectrometer and the NEBULA neutron array. This settled the basis for new proposals, which could address now the multineutron question in the best possible conditions. Incidentally, during our 2012 campaign another experiment was running at RIKEN in the SHARAQ beam line, on the double charge-exchange of a $^4$He target into the 4n system. Their results, published 4 years later, exhibited a 4n signal around the threshold \cite{RIKEN}. 
 This second signal was also weak, but it was obtained in a completely independent way and compatible with ours. We had observed a signal consistent with a 4n state in the range $E(4n)\sim[-1,+2]$~MeV ($2.5\sigma$), while this RIKEN result was consistent with a 4n state at $E(4n)=0.8\pm1.3$~MeV ($4.9\sigma$).

 A ``tetraneutron'' campaign was then planned at RIKEN in 2016/17 along three axes: 1) a new improved experiment on the double charge-exchange of $^4$He into 4n in the SHARAQ beam line; 2) the $\alpha$ removal from $^8$He with a proton target, leading to the formation of the 4n system with little recoil; and 3) the search for the formation of $^7$H states and their decay in flight into 4n. The first two experiments would deduce the 4n energy from all the other charged particles, the third one would measure the neutrons. At this moment, only the second one has led to results, with the strong confirmation of a 4n signal consistent with a 4n state at $E(4n)=2.4\pm0.6$~MeV ($\gg5\sigma$) \cite{Nature}.
 The other two analyses are still in progress, and a new multineutron campaign (4n but also 3n and 6n) is already planned at RIKEN.

\section{Conclusion}

 The first 4n signal was a typical product of basic research, a chain of unexpected events. The idea came from a random read, thanks to the exposure to other fields provided by old libraries. The signal itself was observed during a calibration test of the new probe on data from a previous experiment, while preparing a dedicated proposal. And finally, the series of dedicated experiments over a period of 5 years gave no results due to several issues out of our control that had little to do with physics.
 However, the official version accessible to the general reader is the opposite. A well-thought plan, then a experiment, then the expected result... and then nothing.
 
 It is unfortunate that we almost need to distort what we really do, and keep silent about failures and null results, in order to fit within the standard publication format. The most adverse effect is that we make others believe that this is how our fields work. Funding agencies, but also young researchers, expect well-thought plans, with guaranteed results and no room for the unexpected.
 Another negative effect is the slowing down of research. We know about successes, but we are not aware of when and why some researches did not work.
 It would be beneficial for basic research as a whole to reconsider the way in which we publish, not our results, but our activity.
 
 This tetraneutron result was also a good opportunity to consider outreach and our relationship with the media. We may hope to have total control on the message that will be transmitted to the audience, but we should not expect it.
 Even if most of the media gave a fair treatment to our result, the few that did not left a bitter taste and made us feel against this kind of outreach.
 However, we should keep in mind that publicity on what we do is important, for the society that funds our research but also for the young people that will replace us in the task.
 Finally, we might think that Wikipedia is not a serious source, but many people do use it. If you have not created a page on your results, check if someone else did and verify the content!



\begin{thebibliography}{99}
\bibitem{Review} F.M.~Marqu\'es and J.~Carbonell, Eur.\ Phys.\ J.\ A {\bf57}, 105 (2021).
\bibitem{Nature} M.~Duer \etal, Nature {\bf606}, 678 (2022).
\bibitem{GANIL} F.M.~Marqu\'es \etal, Phys.\ Rev.\ C {\bf65}, 044006 (2002).
\bibitem{Grater99} J.~Gr\"ater \etal, Eur.\ Phys.\ J.\ B {\bf4}, 5 (1999).
\bibitem{Chadwick} J.~Chadwick, Nature {\bf129}, 312 (1932).
\bibitem{Haloes} F.M.~Marqu\'es \etal, Phys.\ Lett.\ B {\bf476}, 219 (2000).
\bibitem{3n} A.~Hemmdan \etal, Phys.\ Rev.\ C {\bf66}, 054001 (2002).
\bibitem{4n} S.C.~Pieper, Phys.\ Rev.\ Lett.\ {\bf90}, 252501 (2003).
\bibitem{Carlos} B.M.~Sherrill, C.A.~Bertulani, Phys.\ Rev.\ C {\bf69}, 027601 (2004).
\bibitem{ArXiV} F.M.~Marqu\'es \etal, arXiv:nucl-ex/0504009.
\bibitem{Jim} J.~Al-Khalili, Nucl.\ Phys.\ News {\bf13}, No.~1, 3 (2003).
\bibitem{RIKEN} K.~Kisamori \etal, Phys.\ Rev.\ Lett.\ {\bf116}, 052501 (2016).
\end{thebibliography}

\end{document}